# Reducing Ohmic Losses in Metamaterials by Geometric Tailoring

**Durdu Ö. Güney,**[1*] **Thomas Koschny,**[1,2] **and Costas M. Soukoulis**[1,2]

[1]*Ames National Laboratory, USDOE and Department of Physics and Astronomy, Iowa State University, Ames, IA 50011*

[2] *Institute of Electronic Structure and Laser, Foundation for Research and Technology Hellas (FORTH), and Department of Materials Science and Technology, University of Crete, 7110 Heraklion, Crete, Greece*

[*]dguney@ameslab.gov

**Abstract:** Losses in metamaterials render the applications of such exotic materials less practical unless an efficient way of reducing them is found. We present two different techniques to reduce ohmic losses at both lower and higher frequencies, based on geometric tailoring of the individual magnetic constituents. We show that an increased radius of curvature, in general, leads to the least losses in metamaterials. Particularly at higher THz frequencies, bulky structures outperform the planar structures.

**Introduction**

Not long after the first experimental verification of negative index of refraction by Shelby *et al.* [1] in 2001, validity of the experiment was questioned. It was claimed [2] that due to the unavoidable losses and dispersion in the negative index metamaterial (NIM) and the geometry used, the transmission measurements would be highly ambiguous for a reliable interpretation of the experiment. One of the authors has demonstrated numerically [3] that it is indeed possible for the NIMs to have good transmission properties, despite the dispersion of their effective permeability and refractive index. However, it has been argued by Dimmock [4] that due to inherent losses, it might be quite challenging with conventional approaches to achieve useful NIM devices in the infrared and visible spectral regions. We [5] studied how the ohmic losses (i.e., total power absorbed by the metamaterial, due to resistive heating per unit cell in the metallic layers of the structure) scale with the size of the split-ring resonator (SRR) structures and found they increase with the decreasing size of the SRRs. More precisely, ohmic losses scale linearly with the resonance frequency for relatively lower frequencies and saturate for higher frequencies before they eventually decrease as the resonance dies.

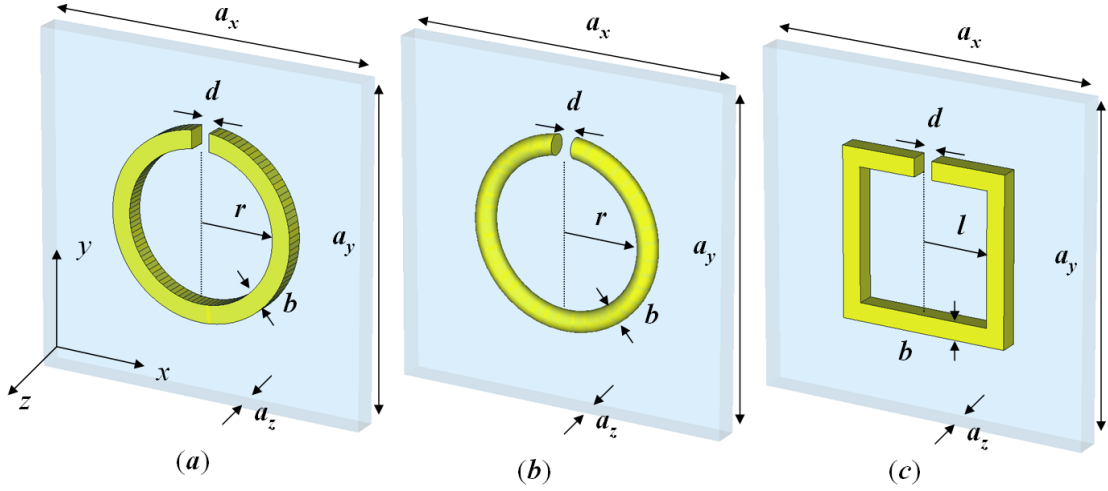

**Fig. 1.** (Color online) The three types of SRRs; metals are described by Drude model for gold [6]. The propagation direction is along the *x*-axis and the SRRs are excited magnetically. (**a**) Circular SRR with a square cross-section, (**b**) circular SRR with a circular cross-section, and (**c**) square SRR with a square cross-section. *d* is the gap width, *b* is the ring thickness, and *r* is the inner ring radius. Unit cell dimensions are such that $a_x=a_y$ and $a_z=b+b_0$, where $b_0$ is the clear separation between adjacent SRRs in the *z*-direction. One side of the square inside the SRR plane in (*c*) is 2*l*.

Ohmic loss is the dominant loss mechanism in metamaterials at substantially high frequencies, the region considered in this paper. Not only the conductivity of the constituent materials leads to ohmic losses, but also the geometric details can have a big influence because of field concentration in resonant structures and non-uniform current distribution.

Reducing the losses is critical to many of the exotic applications expected from NIM technology, including perfect lenses, electromagnetic invisibility cloaks, and others. Here, we discuss two different strategies to reduce ohmic losses in metamaterials at both upper GHz and THz frequencies, respectively. In this region, ohmic loss is usually more problematic than the other loss mechanisms in metamaterials. At lower frequencies (i.e., the upper GHz region) the skin depth is geometrically small compared to the relevant cross-section of the metallic conductors of the metamaterial. The resonant currents are flowing in the metal surface and do not significantly depend on the wire cross-section. Here, we show losses can be reduced by smoothing sharp corners and edges in the current path.

By scaling the metamaterial to higher frequencies (and small unit cells), the skin depth increases relative to the unit cell size. At higher frequencies (THz region) the skin depth is large compared to the unit cell size. Here, we follow a different approach and suggest employment of bulkier wires instead of thin wires.

To achieve our task, we will consider three types of SRRs, common constitutes of metamaterials to provide magnetic resonance, as shown in Fig. 1. In the cases where we compare these three different structures, we will assign the same values to the parameters described by the same symbols in Fig. 1. Our technique should also apply to other metamaterial structures with sharp corners at lower frequencies or consisting of thin wire components at higher frequencies, such as fishnet structures [7] or meandering wires [6].

**Scaling Law and Losses**

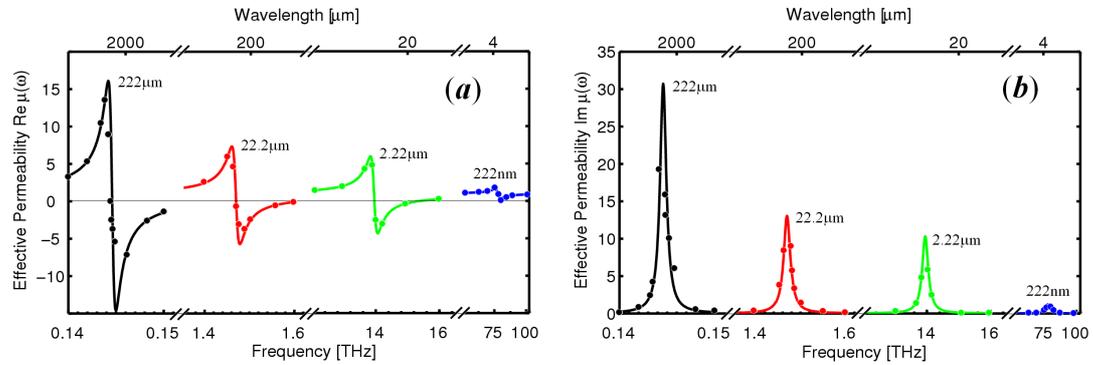

**Fig. 2.** (Color online) Scaling of the (*a*) real and (*b*) imaginary parts of the retrieved effective magnetic permeability with different unit cell sizes *a* of the SRR shown in Fig. 1(*a*), that is *a*=222μm (black), 22.2μm (red), 2.22μm (green), and 222nm (blue). Other parameters are $b' = 12$μm, $r' = 49.8$μm, $d' = 1$μm, and $b_0' = 4.65$μm.

First, we study how the magnetic resonance frequency scales with the unit cell dimensions. For this purpose, we analyze the circular SRR with a square cross-section in Fig. 1(*a*). All simulations are performed using commercial COMSOL Multiphysics simulation software, based on the finite element method. We discretized the computational domain with meshes much finer than the smallest feature size (i.e., gap width, skin depth) of the simulated structures. We imposed periodic conditions on the unit cell boundaries parallel to the direction of propagation and used ports on the perpendicular boundaries to generate the incident field and calculate the *s*-parameters. In our simulations, the ratio of vacuum wavelength to unit cell size ranges between 9—26 at relevant frequencies. Hence, homogeneous effective medium approximation can be used to retrieve the effective [8] constitutive parameters. To limit computational complexity, we only simulate a single unit cell layer of metamaterial in the propagation direction. In the effective medium limit the retrieved effective parameters will apply to longer systems as well.

In Fig. 2, we present the retrieved effective magnetic permeability, $\mu_{\text{eff}}(\omega)$, obtained by the linear scaling of the unit cell dimension, $a$=222μm down to $a$=222nm. The relation between the primed parameter $x'$ and the actual SRR parameter $x$ in Fig. 2 is given by $x = x'10^{-s}$, where s = 0 (black), 1 (red), 2 (green), and 3 (blue). We call s the scaling degree of the SRR. We used the Lorentzian model to fit the simulation data (actual data points are indicated by solid circles in Fig. 2) [9,10]. That is,

$$\mu_{\text{eff}}(\omega) = \mu_\infty - f\omega_0^2/(\omega^2 - \omega_0^2 + j\gamma\omega), \qquad (1)$$

where $f$ is the effective filling factor for the SRR, $\mu_\infty$ is the asymptotic effective permeability far above the resonance frequency $\omega_0$, and $\gamma$ is the dissipation factor. The fitted $\mu_{\text{eff}}(\omega)$ does not go to unity in the limit $\omega\to 0$, because Eq. (1) is only locally valid in the region around the resonance frequency, due to the diamagnetic response [11] of the SRRs. In other words, the diamagnetic response locally shifts the overall Lorentzian response around resonance and therefore Eq. (1) cannot be extrapolated down to $\omega = 0$.

Note, in Fig. 2 the resonance frequency scales up linearly with $1/a$ until the structure size reaches only a few microns, after which the decay of the magnetic response accelerates and linear scaling starts to break down [5,12-14]. An explanation follows. The dependence of the magnetic resonance frequency on $a$ can be estimated as [12]

$$\omega_m = (LC)^{-1/2} = [(L_m+L_e)C]^{-1/2} \approx [(c_m a + c_e/a)c_c a]^{-1/2}, \tag{2}$$

where $L$ and $C$ are the effective inductance and the capacitance of the SRR, respectively, and $c_{m,e,c}$ are constants. In general, $L$ has contributions both from the kinetic energy of the moving electrons and the energy stored in the magnetic field. The former gives the electron self-inductance $L_e \approx c_e/a$, while the latter gives geometric inductance $L_m \approx c_m a$. It is clear from Eq. (2) that for sufficiently large $a$ values (i.e., at low frequencies) the contribution to the effective inductance of the SRR from the kinetic energy of the electrons is negligible and $L$ is predominantly given by the magnetic energy contribution. However, for small values of $a$ (i.e., at high frequencies), the kinetic energy of the electrons becomes comparable with the magnetic energy term and, therefore, starts to add the $L_e$ term to the geometric inductance of the SRR in Eq. (2). This is the main reason why a deviation from the linear scaling of the resonance frequency occurs at higher frequencies. If the structure size is further reduced, the stored energy in the SRR is dominated by the electron kinetic energy. Therefore, the resonance frequency saturates. Although the skin depth effect at a given size $a$ is only roughly considered in Eq. (2), careful analysis shows that the qualitative behavior described above remains the same.

The resonance bandwidth and strength also manifest interesting behaviors in Fig. 2. The strength of the SRR resonance in Eq. (1) is proportional to $f\omega_0/\gamma$. From the fit of our simulation data for $\mu_{\text{eff}}(\omega)$ to the Lorentzian lineshape, the blue curves in Fig. 2 (outside the linear scaling regime) take $f = 0.07$, $\omega_0 = 77.9$THz, and $\gamma = 4.4$THz, so that $f\omega_0/\gamma = 1.24$. Also, when the linear scaling starts to break down, the resonance frequency, $\omega_0$, grows slower than the linear with $1/a$, while $\gamma$ begins to increase rapidly. On the other hand, in the linear scaling regime, black, red, and green curves give $f = 0.167\pm0.006$. This shows that all of these factors—the effective filling factor $f$, the scaling law itself via $\omega_0$, and losses via the dissipation term $\gamma$—are responsible for the rapid decay of the resonance observed outside the linear scaling regime. The decay observed in the linear regime is not as strong. This can partly be attributed to numerical error, because the skin depth relative to structure size becomes extremely small, requiring very fine meshing (especially for the lowest frequency of 151.7GHz). Regarding the scaling of the resonance bandwidth (defined as the full width at half maximum), it is independent from *f* and depends on the ratio $\alpha = \gamma/\omega_0$. Because $\alpha$ is constant in the linear scaling regime, the bandwidth of the SRR resonance also broadens linearly. However, according to our simulation results, $\alpha$ is about 6 times larger for the blue curves, thus giving 6 times broader bandwidth than obtainable if one linearly extrapolates the bandwidth. Therefore, we can argue from $\alpha$ that both the accelerated losses and the scaling law contribute to the broadening of the bandwidth outside the linear scaling regime. This simple analysis of the Lorentzian model clearly shows the breakdown of the linear scaling and the losses act in the same direction. This leads to a rapid decay of the resonance with a simultaneously broader bandwidth than the one expected from the linear extrapolation. However, the relation between the losses and the scaling law, in particular how the breakdown is related with the ohmic losses, requires a more involved analysis.

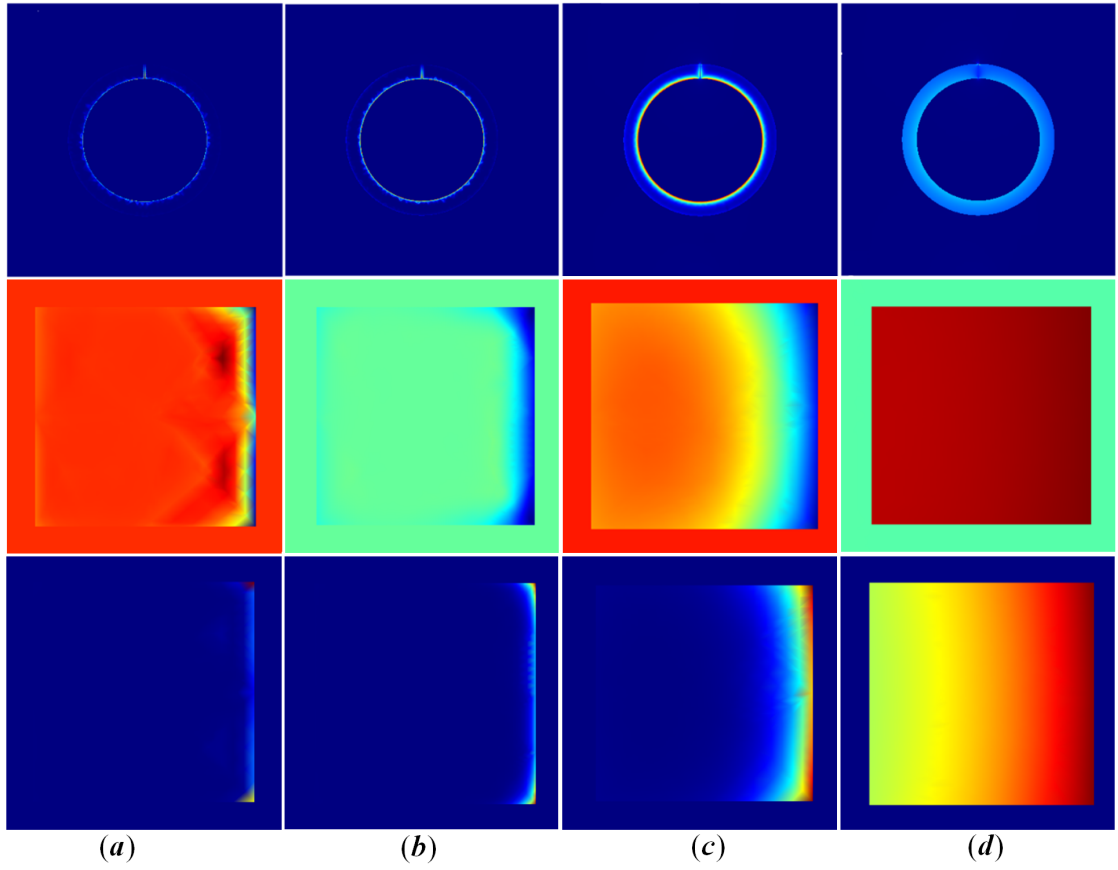

**Fig. 3.** (Color online) Spatial distribution of current density (top and center rows) and power loss density (bottom row) for the SRR of the type described in Fig. 1(*a*) for different scaling degrees (*a*) *s* = 0, (*b*) *s* = 1, (*c*) *s* = 2, and (*d*) *s* = 3. (*a*)-(*c*) correspond to 151.7GHz, 1.527THz, and 14.59THz, respectively, at which $\Re(\mu)$ = -1, while (*d*) corresponds to the resonance at 77.9THz. Cross-sections are obtained by intersecting the SRR with either a central horizontal plane of the SRR (top row) or a central vertical plane (central and bottom rows).

Next, we closely examined the loss characteristics of SRRs at different length scales. We choose the SRRs of the type shown in Fig. 1(*a*) and use the same length scales assumed in Fig. 2. We calculated the time-averaged ohmic loss using three different methods: (i) directly as $P = \int p dV = 1/2 \int \mathbf{E} \cdot \mathbf{J}^* dV$, where $p$ is the power loss density, $\mathbf{E}$ and $\mathbf{J}$ are complex electric field intensity and current density, respectively, (ii) as absorption from transmission and reflection coefficients, and (iii) from the power flow difference between the input and output ports. We determined the ohmic loss obtained in all three methods agree very well.

The columns in Fig. 3 from left to right correspond to decreasing *a* values, leftmost [Fig. 3(a)] being 222μm and rightmost [Fig. 3(*d*)] 222nm. The top row shows the current density amplitude in the central plane of the SRRs. The center row shows the real part of the current density normal to the SRR cross-sections at the central vertical planes. Finally, the bottom row shows the power loss density corresponding to the same cross-sections in the center row. All were calculated at the frequency where $\Re(\mu) = -1$, except for Fig. 3(*d*), which instead displays the resonant distributions, since $\Re(\mu) < 0$ is not achievable [see Fig. 2(*a*)] in this case. Ideally, one could also use a negative index metamaterial and choose the frequency where $\Re(n) = -1$ (when possible) to minimize the impedance mismatch so the calculations would unambiguously highlight the loss behaviors of the structures independent of any reflection. Nonetheless, $\Re(\mu) = -1$ is not only a reasonably good condition to present the idea of how to optimize the losses in this paper, but also avoids additional complications of unfair comparison of loss of structures with different strong magnetic responses. Moreover, it is an interesting region for perfect lens, for example.

At large length scales (i.e., large *a* values), the skin depth ratio is small and most of the current is confined to the edge of the wires [see Figs. 3(*a*)-(*b*)]. However, as the structure size becomes smaller, the skin depth gradually takes over the entire cross-section of the wire and eventually homogenizes the current flow through the wires [see Fig. 3(*d*)]. This occurs because in the linear scaling regime the skin depth scales as $a^{1/2}$, while the structure size scales as $a$, so the skin depth ratio scales as $1/a^{1/2}$. Once the linear scaling breaks down, the decrease of the skin depth is delayed and it takes over the wire cross-section even faster. Geometric dependence of the skin depth may also affect the above scenario, as will be more evident later.

**Losses at Low Frequencies**

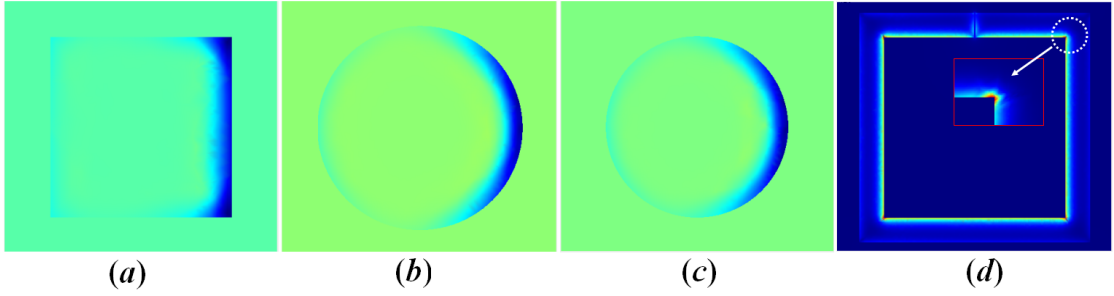

(*a*)            (*b*)            (*c*)            (*d*)

**Fig. 4.** (Color online) The real part of the current density in the cross-sections of different SRRs, which correspond to (*a*) a square cross-section of one side $b$ = 1.2μm [recapped from Fig. 3(*b*)], (*b*) a circular cross-section of diameter $b$ = 1.35μm, (*c*) a circular cross-section of $b$ = 1.2μm, and (*d*) a square SRR with $l'$ = 44.1μm and $s$ = 2. Inset shows the time-averaged ohmic loss at the marked corner. Other parameters of the SRRs in (*a*)-(*c*) are kept the same to not change the resonance frequency significantly and in (*d*) the remaining parameters are linearly scaled down to a second degree (i.e., $s$ = 2) from their original values, which are the same as those of the SRR in Fig. 3(*a*).

At lower frequencies local curvatures in the cross-section perturb the otherwise uniform current distribution inside the wires and contribute negatively to the losses. Therefore, at lower frequencies sharp corners (i.e., large curvatures) should be avoided, where the currents concentrate, such as shown in Figs. 3(*a*)-(*c*).

Our first technique applies to lower frequencies. We select the SRR in Fig. 3(*b*), which has magnetic resonance at 1.527THz, and transform it into another cross-section with smoother corners. The simplest SRR one could design in this way is a circular SRR with a circular cross-section as shown in Fig. 1(*b*). We chose two such SRRs—one with the same cross-sectional area and the other one with the same thickness $b$ (or diameter) as of the original SRR. Figures 4(*a*)-(*c*) display the real part of the calculated current density in these cross-sections at $\Re(\mu) = -1$. The SRR cross-sections in Figs. 4(*a*) and 4(*b*) have the same area, while the diameter of the SRR in Fig. 4(*c*) is equal to one side of the original SRR cross-section in Fig. 4(*a*). As opposed to the current density distribution in Fig. 4(*a*), we observe in Fig. 4(*b*)-(*c*) the current now tends to spread more uniformly over the thin wire layer and gradually diminishes away from the point, where it is the maximum (rightmost point of the circular cross-sections), without concentrating largely at any specific region. This is especially evident if we consider the higher contrast at current density distributions visible in Fig. 4(*a*) along the right edge above certain strength. A more uniform current distribution results in a lower loss, as the power loss density goes with the square of the current density.

**Table 1.** Calculated time-averaged ohmic losses for the SRRs with different sizes and cross-sections. Grey and turquoise shaded regions correspond to a square SRR with a square cross-section and a circular SRR with a circular cross-section, respectively, both obtained by modifying the circular SRR with a square cross-section given in the central row.

| Ring Type | Cross-Section | Scaling Degree (*s*) | Thickness (*b*) | Frequency[*] | Ohmic Loss[*] |
|---|---|---|---|---|---|
| Circular | Square | 0 | 12.00 μm | 151.7 GHz | 6 % |
| Square | Square | 1 | 1.20 μm | 1.522 THz | 15 % |
| Circular | Square | 1 | 1.20 μm | 1.527 THz | 14 % |
| Circular | Circular | 1 | 1.35 μm | 1.648 THz | 12 % |
| Circular | Circular | 1 | 1.20 μm | 1.519 THz | 12 % |
| Circular | Square | 2 | 120 nm | 14.590 THz | 18 % |

[*] Values correspond to Re($\mu$) = -1.

To further investigate the effect of the sharp corners, we also simulated a square SRR similar to Fig. 4(*d*) with *s* =1. [Fig. 4(*d*) corresponds to *s* = 2. Since the SRR has larger skin depth ratio at this scale, it visually makes the effect more pronounced.] We kept the inner area of the SRR the same as that of its circular version [Fig. 3(*b*) or Fig. 4(*a*)] to avoid a significant shift in resonance frequency. Although we found that ohmic loss is even higher, as expected in this case, we calculated that it is only overall 1% larger when compared with its circular version [and 3% larger than the SRRs with circular cross-sections (see Table 1)]. This is reasonable, because the effect of geometric modifications necessary to transform the circular SRR with a square cross-section into a circular SRR with a circular cross-section is more extensive than the effect of the transformation into a square SRR with a square cross-section. In the former, we need to transform all small sharp corners into smoother ones, while in the latter we only have four additional sharp corners, which adds relatively less to the losses.

In Table 1 we summarize the above results. Note that the losses increase as we go down to the smaller dimensions. The frequencies at which $\Re(\mu) = -1$ already comply with the linear scaling. The geometric modifications result in no significant change in the frequencies. For $s = 3$, remember that $\Re(\mu) > 0$ for all frequencies. Therefore, $\Re(\mu)$ not included in the table. Considering a 12% reduction in loss for $s = 0$ with respect to $s = 2$, it should not be surprising why we have only a few percent reduction, due to the different cross-sections (shaded regions). This shows that more powerful approaches to reduce the losses should incorporate a better understanding of the underlying physical mechanisms in the level of electron and phonon dynamics, and possibly the relationship between the scaling law and losses.

**Losses at High Frequencies**

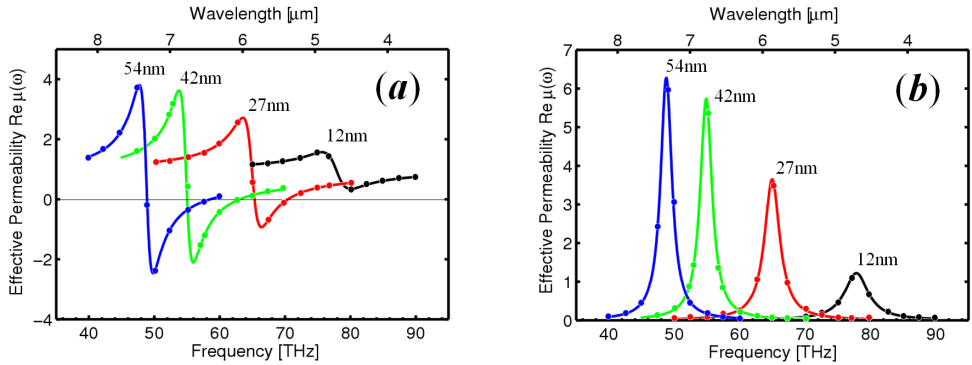

**Fig. 5.** (Color online) (*a*) Real and (*b*) imaginary parts of the retrieved effective magnetic permeability with different ring thicknesses *b* of the SRR, that is *b*=12nm (black), 27nm (red), 42nm (green), and 54nm (blue). Other parameters are correspondingly the same as those of the SRR in Fig. 3(*d*) (i.e., SRR with *b*=12nm, black curve).

After analyzing the low frequency region, now we turn our attention to higher frequencies and select the SRR in Fig. 3(*d*). At this region skin depth ratio is so large that the current density distribution over the cross-section is already uniform. Therefore, no room is left for optimization using the previous technique. Instead, here, we will consider the geometric dependence of the skin depth which is usually overlooked.

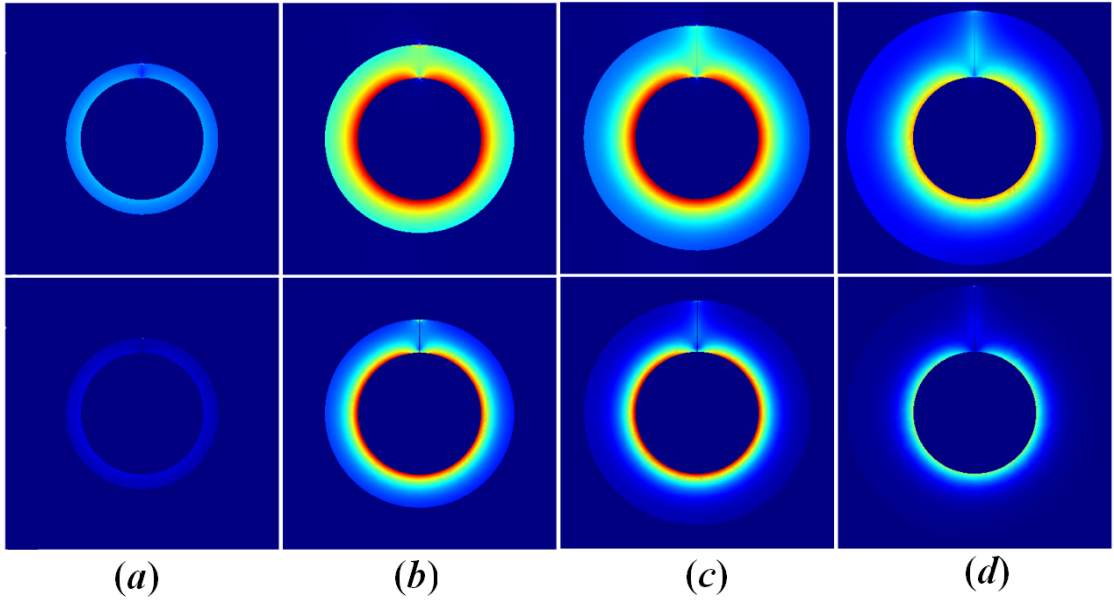

**Fig. 6.** (Color online) Spatial distribution of current density (top row) and power loss density (bottom row) in square-cross-section SRRs employed in Fig. 5 with different ring thicknesses (*a*) *b*=12nm, (*b*) 27nm, (*c*) 42nm, and (*d*) 54nm. Fields are calculated at respective resonance frequencies of the SRRs (77.9THz, 65THz, 55THz, and 49THz, from left to right).

In Fig. 5 we show the retrieved effective permeability for four SRRs with different thicknesses, *b*, ranging from the original 12nm [Fig. 3(*d*)] up to 54nm, keeping the square geometry of the cross-section. Unit cell sizes in the propagation direction are also kept constant; however, $a_z$ is increased accordingly to accommodate the increasing ring thicknesses without changing the separation $b_0$ between neighboring SRRs. We observe that as we increase *b*, initially the strength of the magnetic response linearly increases, but then tends to slow down at high *b* values. Resonant frequency also shifts to lower frequencies by about 37% (mainly due to the increased capacitance). It is remarkable that thicker wires significantly enhance the negative magnetic response, which initially did not reach negative values. One can argue this effect might be simply due to the decreasing resonance frequency (see Fig. 2). However, we checked the magnetic response of the original SRR used in Fig. 2 with a degree of scaling about 2.2. This gives magnetic resonance close to 50 THz. Upon observation this, claim is actually invalid, because the SRR at this frequency gives magnetic responses and losses only in the same order as *b*=27, shown in Fig. 5.

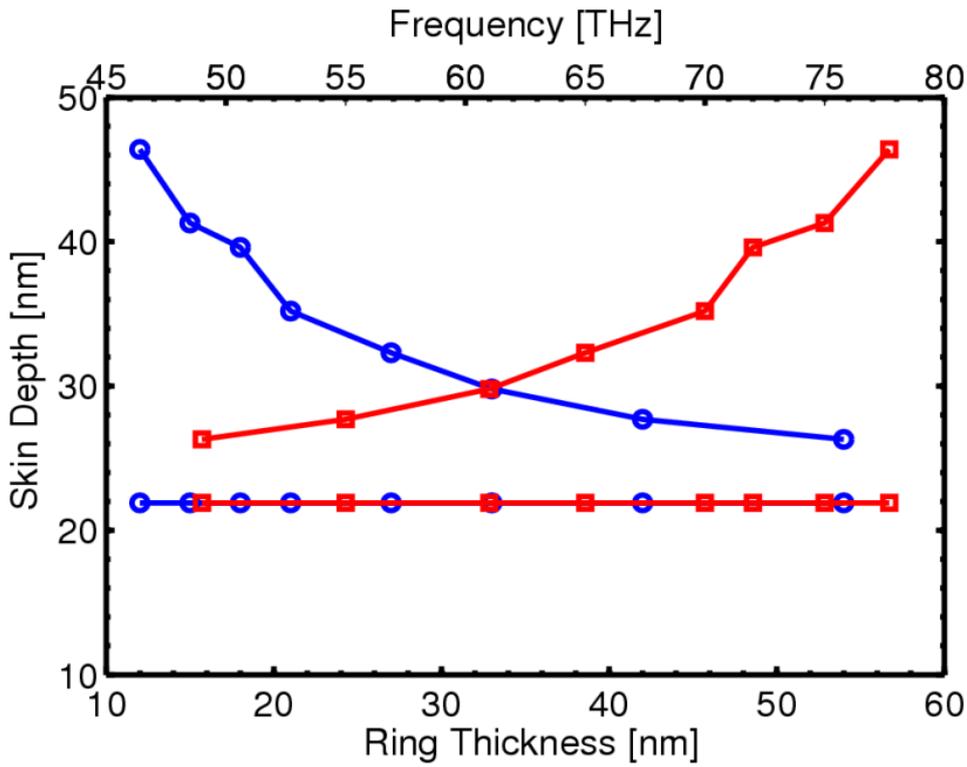

**Fig. 7.** (Color online) Analytical and simulation data for skin depth with respect to ring thicknesses *b* (bottom axis) and corresponding resonance frequencies (top axis). Two lower flat curves are obtained from the analytical (bulk) skin depth formula given in the text at either resonance frequencies of the SRRs (red squares) or corresponding *b* values (blue circles). Similarly, two higher monotonic curves show how the skin depth in simulation changes with resonance frequency (red squares) or ring thicknesses (blue circles).

In Fig. 6 we plot the current density amplitude (top row) and ohmic loss (bottom row) at resonance frequencies of the SRRs employed in Fig. 5 with different ring thicknesses, *b*, ranging from 12nm (left) to 54nm (right). As the SRRs become thicker, the fields become weaker (color scale differs for different panels, average field strength at the innermost surface of the SRRs reduced by about 44% from left to right in Fig. 6). They are effectively expelled from the interior regions of the SRRs, which leaves behind only a thin layer of region where the current actually flows. However, remember the SRRs in Fig. 6 have square cross-sections. This means that although the fields are reduced in the horizontal plane, in fact they are redistributed in the regions closer to the inner edges of the SRRs and contribute more strongly to the magnetic response. [Note *b* for the thickest SRR in Fig. 6(*d*) is 4.5 times larger than that of the thinnest

SRR. In contrast only 44% change results in the average field strength.] As we demonstrate in Fig. 7, this redistribution of the current density also significantly affects skin depth.

For low frequencies, skin depth is dominated by the imaginary part of the dielectric function of the metals and can be approximated as $\delta(\omega) \approx (\sigma\mu_0\omega)^{-1/2}$, where $\sigma$ is the DC conductivity of the metal. For higher frequencies, dielectric function of the metals becomes dominated by its real part and the approximation starts to break down and should be replaced by $\delta(\omega) = (c/\omega)\{1/\Im[\varepsilon(\omega)]^{1/2}\}$. However, both expressions are only valid for semi-infinite metals (bulk skin depth) and can be different for finite metals with arbitrary geometries. In Fig. 7 we show how skin depth changes with the ring thickness, $b$, (blue circles and bottom axis) and the resonance frequency (red squares and top axis) of the SRRs examined in Fig. 6. The two lower flat curves give bulk (analytical) skin depth at either resonant frequency of the SRRs (red squares) or respective ring thicknesses, $b$ (blue circles). Bulk skin depth in the given frequency range is almost constant and about 22nm. However, simulation data (two upper curves), makes it clear that skin depth strongly depends on the underlying geometry; in particular, the thicknesses $b$ of the SRRs. When the wires are thin, skin depth is very large; and, not surprisingly, it approaches the bulk value, as the metal wires become thick and bulky (higher curve with blue circles). It is also worth mentioning that contrary to its bulk value (which, in fact, gradually decreases), geometric skin depth increases with resonance frequency of the SRRs (higher curve with red squares).

The above explained behavior of skin depth has a peculiar impact on ohmic losses. When $b$=12nm, the geometric skin depth is about four times larger than the thickness of the ring and, therefore, fields are almost homogeneous in the cross-section [see Figs. 3(*d*) and 6(*a*)]. For b=54nm, skin depth decreases to 26nm, which is about half the ring thickness. Reduced skin depth, together with larger $b$ values, effectively confines the current to a thin layer of surface, but with a larger cross-sectional area than possible in thin wires. This results in reduced ohmic losses, due to the decreased resistance of the wires.

Table 2 summarizes the calculated losses for different $b$ values at $\Re(\mu) = -1$. We observe that ohmic losses decrease to 11% at $b$=54nm, while at the initially chosen small $b$ values even $\Re(\mu) < 0$ is not achievable. It should be noted that minimum loss obtained here is even less than possible for low frequency SRRs we analyzed at around the 1.5THz range (see Table 1). In Table 2 we also give the fit parameters for the corresponding Lorentzian lineshapes of the SRRs. As expected, the dissipation factor becomes smaller with reduced ohmic losses. In the bulk limit, SRRs also tend to have smaller $\mu_\infty$ and a larger fill ratio *f*. From Eq. (1), the limits of $\omega \to 0$ and $\omega \to \infty$, $\Re(\mu)$ respectively, gives $\mu_\infty + f$ and $\mu_\infty$, respectively, which becomes smaller with increasing *b* values. This shows the diamagnetic response also increases in the bulk limit of the SRRs. Because $\Im(\mu)$ is not affected by $\mu_\infty$, the diamagnetic response gives smaller $\Im(\mu)$ at a fixed $\Re(\mu)$ and, hence, also in part plays a role in reducing the losses.

**Table 2** Calculated time-averaged ohmic losses for the SRRs with different ring thicknesses *b* and corresponding Lorentzian fit parameters $\mu_\infty$, *f* and $\gamma$.

| Thickness (*b*) | $\mu_\infty$ | Fill Ratio (*f*) | Dissipation Factor ($\gamma$) | Frequency[*] | Ohmic Loss[*] |
|---|---|---|---|---|---|
| 27 nm | 0.85 | 0.16 | 2.9 | -- | -- |
| 33 nm | 0.85 | 0.19 | 2.5 | 62.9 THz | 26 % |
| 42 nm | 0.70 | 0.23 | 2.2 | 58.0 THz | 15 % |
| 54 nm | 0.60 | 0.27 | 2.1 | 52.4 THz | 11 % |

[*] Values correspond to Re($\mu$) = -1.

**Conclusions**

In conclusion, we studied the geometric effects on the ohmic losses of the SRRs, a common constitute of negative index metamaterials. We suggested two different strategies to reduce ohmic losses. At low frequencies, losses can be reduced by distributing the available current more uniformly over the SRR cross-section, which can be achieved by increasing the radius of curvature at sharp corners. Interestingly the effect is quite small. At higher frequencies, the current is already uniformly distributed over the cross-section. Therefore, we increased the effective radius of the curvature by going into the bulk limit, which reduces the skin depth dramatically and approaches its bulk value. We observed that in the bulk limit the ohmic losses are appreciably reduced and the magnetic response is significantly enhanced. It may be possible to further reduce losses by additional geometric tailoring. For example, elliptical-cross-section or local geometric tuning of the corners in lower frequencies may also be considered. We also found that at high frequencies the skin depth depends strongly on geometry and the curvature of the conductors. Skin depth actually decreases as the SRR is made more "bulky," partially compensating the reduction of loss arising from the increased conductor cross-section. This dependence of the skin depth needs to be taken into account for studies of metamaterial loss in this frequency regime. Finally, we should also mention there may be more efficient ways of reducing losses in metamaterials by possibly using gain media. However, no demonstration of such technique has been achieved yet.

**Acknowledgments**

Work at Ames Laboratory was supported by the Department of Energy (Basic Energy Sciences) under contract No. DE-AC02-07CH11358. This work was partially supported by the AFOSR under MURI grant (FA9550-06-1-0337), by DARPA (Contract No. MDA-972-01-2-0016), Office of Naval Research (Award No. N00014-07-1-0359) and European Community FET project PHOME (Contract No. 213390).